
\centerline{COMPANIONS OF QSOs AT REDSHIFT 1.1}
\vskip 20pt
\centerline{J.B. Hutchings\footnote{$^1$}{Guest Observer, Canada France Hawaii
telescope, which is operated by NRC of Canada, CNRS of France, and the
University of Hawaii}, D. Crampton$^1$, Andrea Johnson}
\centerline{Dominion Astrophysical Observatory, NRC of Canada}
\centerline{5071 W. Saanich Rd, Victoria, B.C. V8X 4M6, Canada}
\vskip 20pt
\centerline{ABSTRACT}
\vskip 10pt
   We discuss broad- and narrow-band imaging of 7 arcmin fields of 14 QSOs with
redshift $\sim$1.1. The narrow-band filters were chosen to detect
redshifted [O II] 3727A, and the broad bands are R and I, which correspond
to rest wavelengths $\sim$3300A and $\sim$3800A.
In 100 arcsec subfields surrounding the QSOs, we detect an
excess of typically 15 detected objects over the background of 25.
Several of the QSO subfields also
contain an excess of blue (R-I $<$ 1.0) galaxies compared with the
other subfields. Finally, several of the QSO subfields contain an excess of
galaxies with significant narrow-band flux compared with the other subfields,
and many of these are also blue. Most of the QSOs are radio-quiet
in a region of sky overpopulated with z=1.1 QSOs, and 3 others are radio-loud
from other parts of the sky. We suggest that most of these z=1.1
QSOs are in compact groups of starbursting galaxies. In our data, there is no
significant difference between radio-loud and radio-quiet QSOs.
We discuss cosmic evolutionary implications.
\vskip 10pt
    To appear in the Astronomical Journal
\vfill\eject
\centerline{1. INTRODUCTION AND MEASUREMENTS}
\vskip 10pt
   In Hutchings, Crampton and Persram (1993:HCP), we discussed results
from an imaging program to detect companions to QSOs, using broad-band and
also narrow-band (NB) filters tuned
to strong emission lines at the QSO redshift. We described two fields in
some detail at z$\sim$1.1, and initial results from several others. The
principal result was that we find evidence for an excess of faint galaxies
within $\sim$1 arcmin of the QSO, and that among these there is an
excess of blue galaxies and emission-line candidates, compared with other
parts of the 7arcmin fields. The faint galaxy magnitudes are consistent
with bright galaxies at the redshift of the QSO, but their blue colour at
this redshift can only arise if they are starburst objects. The suggestion
from these results is that the QSOs are in compact groups or
clusters of star-forming galaxies. Such a conclusion may have
significant implications for galaxy formation, and cluster evolution.
We also note that early results from the Keck telescope and other recent
investigations (Matthews et al 1994, Soifer et al 1994, Graham et al 1994,
Giavalisco, Steidel and Szalay 1994), are also finding imaging evidence
of clusters around active galaxies at even higher redshifts. Dressler
et al (1993) report on a possible cluster at z$\sim$2.0 from HST imaging.

   We present here the details of the whole sample at z=1.1, including one
further object observed with the High Resolution Camera of the CFHT.
We have 11 radio-quiet QSOs from the region studied by Crampton, Cowley
and Hartwick (1989), and 3 radio-loud QSOs from elsewhere in the sky. The
main program was observed with FOCAM, covering a 7 arcmin field with 0.41
arcsec pixels. The HRCam observations cover a 2.5 arcmin field, with 0.11
arcsec pixels. The images were taken with I and R band filters, and also
a $\sim$75A passband narrow-band (NB) filter at the wavelength of [O II]
3727A at the redshift of the QSO. This dataset allows us to measure
broad-band colours from blue-UV rest wavelengths at the QSO redshift, and
to identify candidates for emission-line companions to the QSO. We can also
make comparisions between the galaxies near the QSO and in the field.
Table 1 lists the objects observed and the data obtained.The data were
 taken in two nights, and towards the end of one of them the
sky brightness increased systematically with time due to scattered
moonlight. There is no vignetting in the fields covered by the CCD.

  Our analysis has three aspects. First, we compare the numbers of objects
in the broad-band images near the QSO and over the field. Second, we
compare the broad-band colours of objects near the QSO and elsewhere.
Finally, we measure NB flux by comparing the NB and I band images, and
compare these near and away from the QSO. In a few fields we obtained
NB images with a filter not matched to the QSO redshift, as an empirical
control. In all cases, we make comparisons between the field near the QSO
and the rest of the image, using the same selection criteria and measurements.
We rely to a lesser extent on absolute values of quantities, counts, and
flux ratios, although we do comment on these and note whether they are
reasonable.

   In making this analysis, there are several basic concerns. These are:
a) a detection criterion for faint objects; b) separation of stars and
galaxies;
c) measurement of magnitudes; d) self-consistency of procedures.
In HCP the signal levels and accuracies of the data were described, but a
more detailed account of the procedure is given here.
The images were first processed with IRAF using standard
procedures to incorporate sky and dome flat-fields and the photometric
standard sequences. The FOCAS software was then used to locate and measure
all objects on the frames. Obvious bright stars -
compact sources $\ge$2 mag
brighter than the QSO, which number about 20 per full field - were also
rejected at this
stage. In many cases, two images were taken with the I filter
to assist in cosmic ray detection, but this was only partially successful
in the automated search, particularly
because the CCD was binned 2x2 (we were concerned
with detection rather than spatial resolution, and the seeing did not warrant
better sampling). In the end we found it necessary to inspect all images
visually to ensure that cosmic rays, CCD cosmetic effects or other artifacts
did not introduce spurious detections. The NB images do not go as deep as the
R and I images, so only objects found by FOCAS on the wide band images were
retained for the statistical analyses. Given the faintness of the objects
and the rather poor sampling/seeing, no attempt was made to differentiate
between faint stars and galaxies, so that the stars are a constant background
for all fields. Thus the final object counts were those found by FOCAS, less
those which are clearly artifacts and those which did not meet the 3$\sigma$
criterion. The fluxes of objects were measured with FOCAS, DAOPHOT, and
the IRAF task Rimexam. Because of variable crowding and image flaws
only the Rimexam results have been used. For this program,
various object apertures and background annuli were tried to
determine optimal values, and noise statistics were derived using
matching apertures on many regions of blank sky. In normal cases the aperture
was 8 arcsec. The entire process was done independently by two of us and
comparison of the results show good agreement in all cases.

The detection limit for faint objects depends on the brightness,
size (i.e. mean signal level per pixel), image quality, and sky
signal. In our data, more compact objects are detected to fainter limits:
in the case of faint galaxies, we reached objects somewhat fainter than 23.
Since the main goal of our work was to determine whether or not
there are excesses of faint galaxies near the quasars, and, if so, what their
overall properties are, restricting the analysis to a sample complete
to some limit is not necessary. The main requirement is one of uniformity
of detection over an entire field. Utmost care was taken in this
regard, and all objects counted and measured are 3$\sigma$ or higher
detections, and our numbers are self-consistent within each frame. We do not
claim
that we are detecting all, or even a known fraction of galaxies present:
we do claim that we can compare the immediate environment of the QSO
with the rest of the frame. Similarly, the colors and NB fluxes can
be compared in the
same internally consistent way.
Thus, \it the comparison of the QSO immediate environment with the
whole frame is based on a self-consistent procedure for each field. \rm

As mentioned above, scattered moonlight affected a portion of the data
and this obviously affects the detectability of faint galaxies. The number
of `background' galaxies counted in the frames correlates well with the sky
brightness, as shown in Figure 1. In the cases of higher sky signal, we
have used Figure 1 to scale the results to intercompare fields. These
corrections are significant for 4 of our 12 z=1.1 fields (5 of 14 QSOs).
Naturally, the object statistics within every image stand on their own in
assessing any excess near the QSO. Our background galaxy counts are
consistent with the those reported by Metcalfe et al (1991).

Our procedure was to catalog all detected objects in the 400 arcsec field,
and to bin them in
16 100 arcsec subfields. In addition, a 100 arcsec subfield
was centred on the QSO. The background counts are taken as the
mean of the subfield counts that do not overlap the QSO subfield (i.e. normally
12). In some instances (typically 1 or less per 400 arcsec field) there
are subfields that have 3$\sigma$ excess counts over the mean. These are
presumed to be distant clusters and are rejected from the background mean
quoted in Table 2. (However, if they are included, the background count is
raised by only 1 and we do not lose the overall
significance of the excess in the QSO subfields. If the high values are
not real clusters, then it follows that about 1 of our QSO subfield excess
counts may also not be real.) The spread of the subfield
source counts is used to estimate the uncertainty in the background value.
Typically, this is within one of the square root of the object counts, so
that the scatter is approximately gaussian. Similarly in the QSO subfields,
the scatter of source counts by different authors or methods is slightly
less than the square root of the number. The significance values of the
excess counts given in Table 2 are derived from these
two uncertainties. Table 2 also includes the results from HCP in which
counts were made on the co-added I and R images. The present work kept the
images separate in order to retain more colour information.

   To compare the galaxies near the QSO and in the field, we measured
the fluxes for all counted galaxies in the QSO subfields and also in one
or more typical background subfields. In the frames free of moonlight,
the photometry is good to $\sim$0.2 mag at I= 23, improving to half that
scatter at 21 mag and brighter.
The NB-I scatter is $\pm$1 mag at I=23, $\pm$0.5 mag at I=22,
$\pm$0.25 at I=20 at the 3$\sigma$ level. The measures were made in the same
way in the NB images. Because of the weaker NB signal, these have faint limits
about 0.7 brighter.

In addition to source counts, broad-band colours and the ratio of NB to
I band fluxes were compared between the background and QSO subfields, to
determine if there are differences. Plots of NB-R with I were used, as
in HCP, to find objects which lie more than 3$\sigma$ away from the
expected measuring scatter: those with high NB fluxes are discussed as
line emission candidates, and numbers compared with the background fields.

With H$_0$=100 and q$_0$ = 0.5 the box size is $\sim$500 Kpc at z=1.1.
We describe below the details of the individual fields, and then
discuss the significance and implications of the results. We describe the
results for the first field in more detail, as the process is similar for all.
\vskip 10pt
\centerline{2. RESULTS FOR INDIVIDUAL FIELDS}
\vskip 10pt

\centerline{1336.8+2848 and 1336.8+2834}
\vskip 10pt

    These two quasars are close together and at very similar redshift (1.124
and 1.113). Thus we could observe them both in the same CCD frame, and use
the same NB filter for both redshifts. The redshifted 3727A wavelengths
are 7916A and 7875A, and the filter centre was at 7906A. The filter passband
has a FWHM of 130A and has a flat peak, so that the z=1.113 object(s) should
have only slightly (if at all) weaker detections than the z=1.124.
In addition, a control filter at 8096A was used on the same field, which
should not detect any QSO-redshift companions.

    The R and I band images were divided into 16 100-arcsec squares and
counts made of faint galaxies in all, as described above. Flux measures were
made on the objects in both QSO subfields and for background comparison on
two other subfields which fill the space between the quasars.

    The results fairly strongly indicate an excess of faint galaxies around
the quasars. This excess consists largely of blue galaxies and objects with
excess flux in the narrow-band passband. These objects are not seen away from
the quasars, and there is no population of emission-line candidates near the
QSOs in the control filter image. We give the details in Table 2 and below.

    In I band, the boxes that do not contain the quasars have an average
of 26 $\pm$ 6
faint galaxies, while the quasar boxes have 45 and 41. The excess of 15
to 19 galaxies is significant at the 2.5 - 3$\sigma$ level. In R band the
figures are similar: background level of 26 $\pm$ 5 and quasar box counts
of 51 and 45: excess of 24 and 19 at the 2.5 - 3.5$\sigma$ level. The narrow
band image counts are lower because of the weaker signal. The background
is 16 $\pm$ 4 and the quasar boxes have 35 and 30 - a similarly significant
excess. The subfield containing the QSO 1336.8+2848 has the higher
counts in all cases.

    In the plots of the R-I colours of objects in the quasar and field boxes
(Fig 2), the field distribution is different at the 95\% confidence level
and the quasar boxes clearly have an excess of blue objects. The spatial
distribution
of the blue objects is shown in Fig 3: 1336.8+2848 is central in its group
while
1336.8+2834 is nearer the edge. Using the criterion of R-I$<$1 for
blue objects, there are 75\% and 64\% in the QSO subfields, compared
with 51\% in the control fields.

    Emission line candidates (Fig 4) differ between the quasar and control
fields, and also between the correct and control NB filter images.
In Figure 4 we show the mean N-I value for a featureless continuum object,
and the 3$\sigma$ scatter from the signal and noise levels in the data. We
regard objects as emission-line candidates if they lie below the 3$\sigma$
curve shown. Thus, the emission line strength required for detection increases
as the object becomes fainter. In the control
filter image, there is no population of emission line candidates anywhere,
although the shorter exposure given means the scatter (shown in Fig 4)
is larger. In the fields away from the quasars, there are 2 emission
line candidates per 100 arcsec box, with the correct filter. In the quasar
boxes with the right filter, there are 5 emission line candidates in
each QSO 100 arcsec box. Of these, many are also very blue objects.
Figure 5 (top) shows a colour-magnitude plot for the QSO and control
subfields. The QSO subfield has more faint blue objects.

    Thus, we conclude that there is an excess of blue \it and \rm emission
line objects in the QSO subfields. The emission line candidates overlap
considerably with the blue objects, and the distribution of emission line
candidates is similar to the blue objects. With so few objects and a sample
that is incomplete at the faint end, it is not possible
to make any statement about spatial structure that might relate to beaming of
radiation from the quasars, or to reddening material. The emission-line
candidates we do have, appear to cluster near the QSOs.

\vskip 10pt
\centerline{0850+140}
\vskip 10pt
    This is a radio-loud QSO. The observed field appears to contain 2 clusters
of faint galaxies. The QSO subfield has a 2.5$\sigma$ excess of 14 faint
galaxies, and a nearby field has an excess of 19 similar objects.
The general background count rate is average at 25. The QSO subfield
has 6 blue galaxies and a control field (not the other
cluster) has 5. The NB filter is centred at 7906A and the redshifted
wavelength for 3727A is 7864A, which is further than most from the NB
passband centre. Apart from the QSO, neither field has any strong
emission-line candidates: the QSO has 1 possible candidate and the control
field has 2. The colour distributions of galaxies are not significantly
different.

   The group of faint galaxies near the QSO is quite compact, with the
QSO at one edge of it (Fig 6). The overall size of this region of faint
galaxies is 30 x 70 arcsec, or 150 x 350 Kpc at the QSO redshift.
\vskip 10pt

\centerline{1335.3+2833}
\vskip 10pt

   This QSO field does not have a clear result. We have 26 objects in the QSO
field and 27 in the control field. Compared with all subfields,
the QSO subfield galaxy counts have an excess
of 5 (1$\sigma$) in I, and of 3 (1$\sigma$) in NB. The R image shows excess
of 9 which is 2.5-3$\sigma$. Visual inspection of the image suggests that if
there is an excess of faint galaxies it is in a larger elongated region
running diagonally through the subfield.

    The colour plots for the QSO and control subfields are almost identical,
and neither one has
anything very blue. The QSO is the bluest object in either field. The next
bluest objects occur equally in both fields and show no spatial placing
of interest. The NB-I plots are also the same for each, and neither has
any good candidates.

    The analysis was repeated for another subfield with fewer galaxies. The
result is very similar, with the same distribution of colours, and much
the same spread of NB/I. In the latter there are slightly fewer objects
on the NB signal side of the spread, but there is nothing which lies
outside the expoected scatter.
\vskip 10pt
\centerline{1336.2+2830}
\vskip 10pt

     The number counts show no excess in the QSO field. The whole field
has high sky brightness and somewhat low galaxy counts at 19 and 15 in
I and R resp. The QSO does have a group of very faint galaxies while most
other fields have brighter ones (Fig 6). There is a definite cluster
of brighter faint galaxies (37 and 30)
in one box away from the QSO. The QSO box has one blue galaxy and an average
colour slightly bluer than the control fields. There are no blue galaxies
in the control fields (which do not include the rich cluster).

    There is one emission line candidate in the QSO field and none in
the control fields. Of the galaxies in the QSO field, 5 of 11 lie
on the `emission' line side of the plot and none on the other: the control
fields galaxies are evenly spread.

   Overall, there is marginal evidence for a sparse, faint group
associated with the QSO. The higher than average sky brightness causes
the formal galaxy count excess to be zero, however.
\vskip 10pt

\centerline{1337.4+2744}
\vskip 10pt

    This QSO has a small excess of galaxies in its box (4 or 1$\sigma$)
above the mean of 17 in I and 13 in R. The NB image has an excess (also
1$\sigma$) of 2 above a mean of 8. The immediate vicinity of the QSO does not
have a small group of galaxies. A correction for higher sky brightness
had to be applied for this field, so the faintest galaxies will be missed.

    However, the colours of the nearby galaxies are bluer than in the
control field. There are 5 (of 21) galaxies bluer than the QSO, all
of them faint. These are grouped in a line extending either side of the
QSO (Fig 6). The control field has no blue galaxies at all.

    The NB ratios show 3 emission line candidates (of 10, including the
QSO: Fig 5), two of which are
among the blue ones (the other 3 blue ones have no NB measure). These lie
along the line defined by the blue galaxies. The
control field has no emission line candidates in 4 measurable galaxies.

     The results are consistent with a sparse group of faint blue galaxies
around the QSO along a line on either side (Fig 5). The correction for sky
brightness increases the excess count from 4 to 7 (Table 2).
\vskip 10pt

\centerline{1339.5+2738 and 1339.8+2741}
\vskip 10pt

    These quasars have redshift of 1.175 and 1.185 respectively,
and are in the same CCD field. They were observed with the 8096A filter,
and the redshifted 3727A wavelengths are 8106A and 8143A. 1339.8+2741
is further
from the filter bandpass centre but is within the filter FWHM (8034-8159A).

   The results in this field are not as clear, and suffer from high sky
brightness. The faint galaxy counts
in the two QSO fields are 18 and 17 in R and 18 and 20 in I, compared
with 11 $\pm$ 2 in R and 11 $\pm$ 3 in I, in typical subfields.
Thus there is an excess of about 7
in each QSO field, which has 2-3$\sigma$ significance.
While the I and R bands are not totally independent, we note that a similar
excess is found in both. The counts in the NB
images similarly give an excess of 2$\sigma$ and 3$\sigma$ in the QSO fields.
The correction for sky brightness raises the excess to about 20 for each
of these QSOs. We conclude that the QSOs are at least in small groups of
galaxies. The distribution of the observed
galaxies around the QSOs is tight in these cases too, with several lying
within ~10 arcsec radius (50 Kpc at the QSO).

   The colours of the galaxies in the QSO and control fields are not
different, and there are few blue objects (R-I$<$-1). The QSOs are both
blue; 1339.5+2738 has two others and 1339.8+2741 has one other in their
fields. A typical field has 3 blue galaxies.

   The plots of I vs NB-I are similar for all subfields, with no candidates
for emission-lines, including the QSOs themselves (Fig 5).
The scatter of points
is symmetrical as we would expect for no emission line sources, and lies
nicely within the expected scatter for a null result.
\vskip 10pt

\centerline{1632+391}
\vskip 10pt

   This was observed on a later run, with the same filters and HR Cam.
The field is smaller, the sampling different, and the exposures a little
different. The field was divided into 4 sections, one surrounding the QSO.
The galaxy counts were 18 for the QSO subfield and 20 in the diagonally
opposite field,
of generally fainter galaxies. The other two subfields have 6 and 8 galaxies.
The QSO field
has a tight group of irregular galaxies near the QSO, and also has
an excess of blue galaxies grouped near the QSO.

   There are no strong emission line candidates in either the QSO or control
subfields. Of 6 moderate emission-line candidates, 3 are close to the QSO.
Only one is blue. Of the 11 nearest
neighbours of the QSO, 6 are as blue as or bluer than the QSO, and 3 others
are emission-line candidates.

   In this field, the data suggest a compact blue cluster associated with
the QSO, although a wider overall field would have been useful to improve the
galaxy count statistics. The better sampling and resolution of HR Cam however,
allow us to note the structure and size of the QSO companions. The galaxies
are shown in Figures 7 and 8. They are all irregular-shaped and have
luminosity scale lengths of 4-5 Kpc at redshift 1.1. Their overall diameters
to 1\% of the sky ($\sim$26 mag/arcsec$^2$) are about 13 Kpc at redshift 1.1.
Many of them also have low central brightness.
Thus these are not symmetrical and standard galaxies.
\vskip 10pt
\centerline{1335.2+2685, 1336.2+2689, 1339.4+2756}
\vskip 10pt
 These QSO fields were summarized in HCP, and Table 2 gives the measured
quantities for them. They all contain a significant excess of faint galaxies
in the QSO subfield, and of blue galaxies and emission line candidates.
The sky signal is low and the faint galaxy detections are good. Figure 6 shows
the distribution of blue and emission-line candidate galaxies in the
QSO subfields. In all these we do not see a clear area of higher galaxy counts
but there are usually several faint galaxies close to the QSO, and
statistically
the whole subfields have an excess of blue and emission-line objects. These
counts reach the faintest limits in the dataset: applying brighter
detection limits reduces the numbers but does not alter the overall excesses
we find in the deeper sample.

\vskip 10pt
\centerline{3. DISCUSSION}
\vskip 10pt
  In several instances (9 of 14) we see a significant (3$\sigma$)
excess of galaxies associated spatially with the QSO.
In most individual cases, and cumulatively, the QSO environments contain
a small excess of blue galaxies (R-I$<$1) and emission-line candidates
compared with the field. In all fields combined, there is an excess of
155 objects in the QSO subfields compared with the other subfields, and this
is significant at $>5\sigma$ level.

While the field is known to contain faint blue
galaxies at lower redshift, the association of galaxes this blue with
the QSO at z=1.1 implies a very young population. Further, there are several
more emission-line candidates near the QSO than in the field, also suggesting
a population with active star-formation. The combined weight of these
independent indicators suggests that several of the QSOs in our
program do lie in groups of what may be very young galaxies.

   At a redshift of 1.1 for H$_0$=100 and q$_0$=0.5 the distance modulus is
43.7. The k-corrections for an old population are 0.8mag in I and 1.5mag in R
(see also Metcalfe et al 1991; Coleman, Wu, and Weedman 1980). As in HCP, we
calculate for a starburst population, these are -1.9mag and -1.1mag
respectively.
Thus, a I=22 or R=23 galaxy at this redshift has absolute magnitude in these
bands of -22.4 for an old population; -19.6 for a starburst; and -20.9
for an E+A population. Objects at this redshift with R-I$<$1 must
be starbursts. If we are observing a real population of galaxies
associated with the QSOs, the blue ones are star-forming galaxies which
may be of luminosity comparable with the LMC - i.e. of moderate to low
mass. We note that the negative k-correction for very young stellar
populations at this redshift make such galaxies brighter by up to 2
magnitudes: the colours and brightness thus indicate strong evolution of
such QSO companion galaxies.

The QSO-associated groups are compact
and the galaxies are small. It is possible that we are seeing galaxy
formation in small groups at this redshift. At low ($<$0.5) redshift,
radio-loud QSOs lie in larger and fairly rich clusters
(e.g. Yee and Ellingson 1993), while radio-quiet
QSO have no more than 6 or 8 companions, and often fewer.
Thus, the fact that groups such as we see at z=1.1
are not seen around QSOs at low redshift indicates that these clusters
evolve significantly in the time interval since z=1.1. The elapsed time
from z=1.1 to 0.5 is about 2.5 Gy, which considerably exceeds the lifetime
for an individual QSO. Thus, the different environment of lower redshift QSOs
implies that the site of QSO
activity changes in this interval. In addition, the compact nature of
the clusters, the small size and low mass of the galaxies themselves, may
mean that they fade and perhaps merge into fewer galaxies which would be
seen as small groups at lower redshift epochs, if we were to come across
them. We also note that the evolution of a starburst would cause fading
by 1 - 2 visible magnitudes in a relatively short time, and faster than
the k-correction changes with decreasing redshift. Thus, they would
become less easily visible as they evolve.

    As an empirical experiment, we artificially redshifted the image
of a z=0.2 cluster (Abell 2390) to z=1.1. This was done by altering the
sky signal and noise to mimic the redshift-fading, and changing the pixel
scale appropriately. The redshifted z=0.2 cluster did not resemble the
groups we see here: at z=1.1, A2390 is considerably larger and less centrally
concentrated. The brightest galaxies also appear larger and more
regular. It is noticeable how many faint galaxies simply disappear: it
is clearly desirable to have fainter detection limits and dark sky
for further work. Nevertheless, it seems clear we are not seeing the
equivalent of z=0.2 Abell clusters, at z=1.1.

      There is no large difference between the radio-loud and radio-quiet
QSO environments in our small sample. However, the radio-loud QSOs do
appear to be in richer cluster environments on average, than the radio-quiet.
Disregarding (or correcting) the high background fields, the excess counts
in the QSO subfields range from 7 to 25, with one subfield showing no
measured excess.
The colours of the faint galaxies in the QSO subfields and the numbers
of emission-line candidates generally support the conclusion that the
QSOs are in groups or clusters of star-forming galaxies. These groups are
compact compared with recent epoch clusters and the galaxies themselves
are irregular and compact.

   In conclusion, then, it appears that the environment of many, if not all,
QSOs may evolve significantly between
redshift 1.1 and 0.6 and lower. Since the lifetime of individual QSOs is
considered to be less than the time between these epochs (from considerations
of the radio structures, for example), this means that the site of QSO
activity changes with epoch. In addition, the clusters we see at redshift
1.1 will evolve passively and by merging, and will not be seen in this form
in the present day universe. Clearly, it is of value to study the
QSO environment at other redshifts, and also to obtain spectra of the z=1.1
companions to understand their composition, state of star-formation, and
dynamics.

\vfill\eject
\centerline{\bf References}
\vskip 10pt
Coleman G.D., Wu C-C, Weedman D.W., 1980, ApJS, 43, 393

Crampton D., Cowley A.P., Hartwisk F.D.A. 1989, ApJ, 345, 59

Dressler A. et al 1993, ApJ, 404, L45

Giavalisco M., Steidel C.C., Szalay A.S., ApJ, 425, L5

Graham J.R. et al 1994, ApJ, 420, L5

Hutchings J.B., Crampton D., Persram D. 1993 AJ, 106, 1324 (HCP)

Metthews K. et al 1994, ApJ, 420, L13

Metcalfe N., Shanks T., Fong R., Jones L.R., 1991, MNRAS, 249, 498

Soifer B.T., et al 1994, ApJ, 420, L1

Yee H.K.C. and Ellingson E. 1993, ApJ, 411, 43

\vfill
\centerline{\bf Figure captions}

1. The number of faint galaxies per 100 arcsec subfield in R and I bands, as a
function of sky signal level. Dashed curve is function used for correction.

2. Histograms of R-I for faint galaxies in QSO subfields and control
subfield from Figure 3. The QSO subfields have a larger fraction of blue
(R-I$<$1) galaxies.

3. Distribution of galaxies in subfields in double QSO field. Circles have
radius 1 arcmin centred on QSOs.

4. NB colour-magnitude diagrams for double QSO subfields. The horizontal line
is the value for flat continuum with no emission line, and the curved
lines are the 3$\sigma$ scatter expected for the images. Emission-line
candidates lie below the lower curved line. The QSOs are marked as
asterisks. The lower panel is from both QSO subfields combined, with the
control NB filter.

5. Top: colour-magnitude diagram for double QSO field. QSO subfields have
more faint blue galaxies than control subfields. Lower: NB colour-magnitude
plots as in Figure 4.

6. The distribution of faint objects in 6 QSO subfields, showing blue
and emission-line candidate objects.

7. Galaxies in the neighbourhood of 1632+391, taken with HR Cam and 0.6
arcsec seeing. The QSO is the brightest object.

8. Contour plots of galaxies close to 1632+391. All objects shown are
significant detections. Note the irregular and compact shapes.
The QSO is the bright object in the upper right.

\vfill\eject

\centerline{Table 2. Galaxy count statistics}
\vskip 10pt
\settabs 12\columns
\+Name &&~~z &~~~~~R &&~~~~~I &&~~~NB  &~~~~Corr XS$^b$ &&~\#Blue$^c$
&~~~~~\#Emis$^d$ \cr
\vskip 3pt
\hrule
\vskip 5pt
\+0834+250 &&1.122* &&50-31=19(3$\sigma$)$^a$ &&&&&18 &17/55=.31 (1) &&~18\cr
\+~~~field &&&&&&&&&&6/21=.29 &&~~4\cr
\vskip 4pt
\+0850+140 &&1.110* &&39-25=14(3$\sigma$) &&&&&17 &9/22=.41 (1) &&~~0\cr
\+~~~field &&&&&&&&&&5/13=.38 &&~~0\cr
\vskip 4pt
\+1335.2+2685 &&1.090 &&43-31=12(2$\sigma$) &&&&&12 &33/48=.69 (8) &&~~4\cr
\+~~~field &&&&&&&&&&10/19=.53 &&~~0\cr
\vskip 4pt
\+1335.3+2833 &&1.124 &24-15=9(3$\sigma$) &&26-21=5(1$\sigma$)
&&16-13-3(1$\sigma$) &&12 &18/24=.75 (0) &&~~0\cr
\+~~~field &&&&&&&&&&12/15=.80 &&~~0\cr
\vskip 4pt
\+1335.8+2834 &&1.086 &&43-25=17(3$\sigma$) &&&&&20 &16/33=.48 (3) &&~~8\cr
\+~~~field &&&&&&&&&&13/34=.38 &&~~0\cr
\vskip 4pt
\+1336.2+2689 &&1.088 &&40-30=10(3$\sigma$) &&&&&10 &15/23=.65 (6) &&~~4\cr
\+~~~field &&&&&&&&&&6/15=.40 &&~~3\cr
\vskip 4pt
\+1336.2+2830 &&1.116 &13-15=-2 &&19-19=0 &&10-10=0 &&~0 &11/13=.85 (0)
&&~~1\cr
\+~~~field &&&&&&&&&&18/22=.82 &&~~0\cr
\vskip 4pt
\+1336.8+2848 &&1.124 &51-26=25(3$\sigma$) &&45-26=19(3$\sigma$)
&&35-15=20(3$\sigma$) &&25 &27/36=.75 (9) &&~~5\cr
\+1336.8+2834 &&1.113 &45-26=19(3$\sigma$) &&41-26=15(2$\sigma$)
&&34-15=19(3$\sigma$) &&20 &21/33=.64 (4) &&~~5\cr
\+~~~field &&&&&&&&&&24/47=.51 &&~~2\cr
\vskip 4pt
\+1337.4+2744 &&1.120 &16-13=3(1$\sigma$) &&21-17=4(1$\sigma$)
&&10-8=2(1$\sigma$) &&~7 &11/16=.69 (5) &&~~3\cr
\+~~~field &&&&&&&&&&~5/13=.38 &&~~0\cr
\vskip 4pt
\+1339.4+2756 &&1.095 &&41-33=8(3$\sigma$)  &&&&&~7 &12/24=.50 &&~~5\cr
\vskip 4pt
\+1339.5+2738 &&1.175 &18-11=7(2$\sigma$) &&18-11=7(2$\sigma$)
&&13-7=6(2$\sigma$) &&19 &18/31=.58 (-3) &&~~0\cr
\+1339.8+2741 &&1.185 &17=11=6(2$\sigma$) &&20-11=9(2$\sigma$)
&&16-7(3$\sigma$) &&20\cr
\+~~~field &&&&&&&&&&12/18=.67\cr
\vskip 4pt
\+1632+391 &&1.082* &21-9=12(3$\sigma$) &&18-7=11(3$\sigma$)
&&9-6=3(1$\sigma$) &&(28) &14/21=.67 (4) &&~~3\cr
\+~~~field &&&&&&&&&&~6/13=.46 &&~~1\cr
\vskip 3pt
\hrule
\vskip 10pt
* = Radio-loud

$^a$ Numbers in this column based on combined R and I images.

$^b$ Galaxy excess corrected for sky brightness to background count rate of 30
per subfield

$^c$ Number of galaxies with R-I$<$1 / total in subfield,
not corrected for sky brightness. (Numbers in perentheses are excess of
blue objects in QSO subfield.)

$^d$ Emission-line candidate objects outside 3$\sigma$ scatter,
not corrected for sky brightness.
\end